\documentclass{article}

\usepackage{arxiv}

\usepackage[utf8]{inputenc} 
\usepackage[T1]{fontenc}    
\usepackage{hyperref}       
\usepackage{url}            
\usepackage{booktabs}       
\usepackage{amsfonts}       
\usepackage{nicefrac}       
\usepackage{microtype}      
\usepackage{lipsum}
\usepackage{graphicx}
\graphicspath{ {./images/} }

\title{Beyond Privacy Regulations: An Ethical Approach to Data Usage in Transportation}

\author{
 Johannes M. van Hulst \\
  TomTom International BV\\
  Amsterdam, Netherlands \\
  \texttt{mick.vanhulst@tomtom.com} \\
   \And
 Mattia Zeni \\
  TomTom International BV\\
  Amsterdam, Netherlands \\
  \texttt{mattia.zeni@tomtom.com} \\
  \And
 Alexander Kröller \\
  TomTom Location Technology Germany GmbH\\
  Berlin, Germany \\
  \texttt{alexander.kroeller@tomtom.com} \\
   \And
   Cassandra Moons \\
  TomTom International BV\\
  Amsterdam, Netherlands \\
  \texttt{cassandra.moons@tomtom.com} \\
   \And
   Pierluigi Casale \\
  TomTom International BV\\
  Amsterdam, Netherlands \\
  \texttt{pierluigi.casale@tomtom.com} \\
}

\begin{document}
\maketitle
\begin{abstract}
With the exponential advancement of business technology in recent years, data-driven decision making has become the core of most industries. With the rise of new privacy regulations such as the General Data Protection Regulation in the European Union and the California Consumer Privacy Act in the United States, companies dealing with personal data had to conform to these changes and adapt their processes accordingly. This obviously included the transportation industry with their use of location data. At the other side of the spectrum, users still expect a form of personalization, without having to compromise on their privacy. For this reason, companies across the industries started applying privacy-enhancing or - preserving technologies at scale in their products as a competitive advantage. In this paper, we describe how Federated Machine Learning can be applied to the transportation sector. We present use-cases for which Federated Learning is beneficial in transportation and the new product lifecycle that is required for using such a technology. We see Federated Learning as a method that enables us to process privacy- sensitive data, while respecting customer’s privacy and one that guides us beyond privacy-regulations and into the world of ethical data-usage.
\end{abstract}


\section{Introduction}
Data is transforming the way people do business and how they make decisions in industries, including transportation. Modern connected cars are equipped with complex electronic systems that collect data from a multitude of sensors and use it to, among others, improve the driving experience, to reduce fuel consumption and to provide smart maintenance solutions. This market segment is facing an enormous growth, reaching 65 million connected cars in 2019\footnote{https://www.statista.com/topics/1918/connected-cars}. At the same time, people are interested in a personal user experience in the products they use, also in the automotive sector\footnote{https://martechseries.com/mts-insights/guest-authors/the-connected-car-is-poised-for-acceleration-and-hyper-personalization}. The available data then should be used to answer specific users’ needs with tailored applications, services and use-cases. But to do so, companies have to study and understand users’ preferences and behaviour so that to be ready to react to them in the best way possible. Data is considered the new oil \cite{humby2006data}, enabling new opportunities for advanced analytics (e.g., personalization) like never before, but this does not come without cost. By collecting user information, companies have to guarantee that such data is handled according to the latest privacy regulations. And different types of data require different measures: those generated in the healthcare sector for example are classified under a special category \cite{regulation2016regulation} because considered sensitive from a privacy point of view. In the transportation sector, the most sensitive private information is the user’s location. General Data Protection Regulation (GDPR) is the first regulatory initiative to explicitly mention location data in the context of privacy-sensitive information, as from the location of an individual one is able to infer highly privacy-sensitive information. A study performed by de Montjoye et al. \cite{de2013unique} showed that in a dataset where the location of an individual was specified hourly, four spatio-temporal points (e.g. GPS location) were sufficient to identify 95\% of the individuals.
Such findings gave rise to the dawn of privacy-related regulations and led companies to change their course by anonymizing and/or removing their data at the request of the customer. However, even anonymizing data does not guarantee that the privacy of the user cannot be breached. An example is a study, performed by Rocher et al. \cite{rocher2019estimating}, that found that using a particular model and 15 demographic attributes, originating from heavily sampled anonymized datasets, was sufficient to correctly identify 99.98\% of Americans. An example would be the identification of an anonymized sample as John Doe, a 50-year old male from Berkely with lung cancer. This shows that even when a company is compliant with all privacy-related regulations, it cannot guarantee the customer’s privacy. As such, we ask the reader to not just read this paper from the current state-of-affairs, but also from the perspective where these regulations should be a driving force to do better. Here we align our thoughts with what is mentioned in the EU AI Guidelines\footnote{https://www.coe.int/en/web/artificial-intelligence/-/new-guidelines-on-artificial-intelligence-and-data-protection}: “\textit{manufacturers and service providers should assess the possible adverse consequences of data used in AI applications on human rights and fundamental freedoms, and, considering these consequences, adopt a precautionary approach based on appropriate risk prevention and mitigation measures}”. This does not solely entail privacy regulations, but also encourages ethical use of data. Many companies have already started addressing this problem by applying so-called \textit{privacy enhancing technologies (PET)}, which are not mandatory as per the current privacy regulations but are considered the next step towards a more secure and ethical use of data.

With this paper we want to highlight how the transportation sector could deal with users’ data in a privacy compliant and ethical way and how this can be exploited as a competitive advantage on the market. We present where a specific PET technology, Federated Machine Learning, can be used in the transportation sector, affecting the lifecycles and which benefits it brings to our users.

\section{Privacy-enhancing technologies}
\label{sec:headings}

Privacy-enhancing technologies (PET) are methods that allow for processing privacy-sensitive data in a secure manner with the goal of protecting users their personal identifiable information. PETs are an element of the privacy-by-design paradigm, that intends to create systems and processes that at their cores mitigate the privacy risks, without compromising the quality of the data-driven decisions. This is relevant in scenarios where advanced technologies can reverse the process of anonymized data with the goal of obtaining privacy-sensitive information \cite{de2013unique}. Currently, but most importantly in the future, it will be hard to guarantee the anonymity of data, even for privacy compliant organizations. Applying PETs is considered a competitive advantage for companies and the possibilities that PETs enable are big, especially in relation to Artificial Intelligence.

Based on the efficacy and strength of the various PET solutions, we distinguish between two main categories: \textit{soft} and \textit{hard}. The former commonly refers to improving the processes and how people deal with data, such as \textit{assuming trusted data processors, assuring compliances to the regulations and access control}. The latter refers to technical solutions based on mathematical principles that create formal guarantees. Examples of hard PETs are \textit{secure multi-party computation, zero-knowledge proof, and full homomorphic encryption}. An important notion is that these solutions can be combined to increase their overall efficacy\footnote{https://royalsociety.org/-/media/policy/projects/privacy-enhancing-technologies/privacy-enhancing-technologies-report.pdf}.

Among the different techniques, one that is gaining increased interest by both the scientific community and the industry is \textit{Federated Machine Learning} \cite{konevcny2016federated, yang2019federated}. This approach allows for the execution of Machine Learning tasks without the centralization of data. Instead, data is stored on the devices who generated them. The potential of such technology is huge, and many companies are starting to adopt it, creating their own implementations and devising interesting use-cases. Among them are Google \cite{mcmahan2016federated} and IBM \cite{verma2018federated}, who both integrated this method in their products. Our view is that Federated Machine Learning might become the default standard in the coming years and consequently the decision was made to start researching this subject and to start integrating this technology into our products.

\subsection{Federated Machine Learning}

In classical data analysis solutions that may or may not use machine learning, all the data is moved to a centralised server. With the rise of privacy-regulations, companies are forced to anonymise privacy- sensitive data and have to delete them if the customer requests to do so. Without the underlying data, performing advanced analytics in a classical way becomes impossible. An alternative to centralising the data of the customers is to utilize a distributed solution, such as \textit{Federated Machine Learning}.

Federated Machine Learning is considered a privacy-aware machine learning method, where devices that collect data, perform computations locally and send only the aggregated results to a central entity. As the computations are performed locally, the privacy-sensitive data does not leave the device, making it useful for use-cases where customers do not want to share their data, or where uploading the data is unfeasible for technical reasons. From a company’s perspective, the benefits of using this technology can be summarized as follows: (1) being proactive with respect to possible changes regarding privacy- regulations, (2) use it as a competitive advantage for ethical usage of privacy-sensitive data, (3) empower personalization and (4) reducing the costs of the cloud infrastructure.

Federated Machine Learning is not always the best fit for a data-driven product. If the to-be-processed data is not privacy sensitive or unfeasible to move to a central location, i.e., due to their size, it is not worth going through the extra steps and lifecycle changes that come with Federated Machine Learning. Additionally, Federated Machine Learning is currently not applicable to all types of problems. It is very effective in situations where the \textit{user’s reaction} to an action generated by the system is used directly in the learning process. By reaction we refer to the pressing of a confirmation button, a label\footnote{In supervised learning, the machine learning task is based on examples of input-output pairs. In the context of this paper, the input can be described as \textit{the location data} collected from the GPS sensor of the car, while the output is the \textit{label} that the user assigns to those locations, like “Home”. A supervised learning algorithm analyzes the training data and produces an inferred function, which can be used for mapping new inputs to outputs.} assigned to a predicted location, or simply the fact that the user accepts a suggested detour.

\section{Applications for the Transportation Sector}
In this section we present various use-cases that are of interest for intelligent transportation systems. Ones that can be addressed by utilizing Federated Machine Learning. We collected the individual use-cases through interviews with the stakeholders in the product units of our company.

We would like to emphasize that these use-cases can be tackled with classical machine learning solutions, meaning that Federated Machine Learning is not mandatory in these settings. However, we encourage the reader to take into account the changing landscape, both from a legal and an ethical perspective. First of, even though customers currently agree to centralizing their data, one cannot assume that, overtime, this remains unchanged. Second of, privacy regulators are now ready to enforce the regulations, thanks to the insights that were gained in the recent past.

\subsection{Business-to-Consumer}
This section describes use-cases in the Business-to-Consumer market segment of the transportation industry where Federated Machine Learning can be applied. We describe services that may benefit drivers in their daily usage of smartphones applications or in-car navigation systems.
\\

\textit{Destination prediction} is described as the process of predicting automatically the destination(s) of a driver without additional manual tasks, e.g., by typing on the screen or using voice commands. The process of destination prediction includes the usage of historic user’s behaviour and other features such as time of the day and weather. In the canonical way, the process works by moving the user’s trips history to a backend server and then infers patterns by using Machine Learning algorithms. In the federated setting, one does not have to move the privacy-sensitive data, such as location data, to a centralized server. In conjunction with other encryption methods, it mitigates risk by solely transmitting (encrypted) models to a centralized server. At the same time, it does not compromise on the traditional benefits of Machine Learning, such as the ability to infer knowledge of one user’s patterns to improve the experience of another.
\\

An extension of destination prediction is \textit{route prediction}. The goal is to provide a route from to a destination while taking into consideration the personal preferences of the customers. This differs from the traditional setting where users are suggested identical routes that base its computations on parameters such as time (fastest route) or distance (shortest route). A customer could prefer to drive a little longer to avoid traffic jams or has a preferred route for her usual commute Home – Work.
\\

Another key aspect in a navigation system is the \textit{Estimated Time of Arrival (ETA)}. This is described as the time at which the driver reaches his or her destination. The calculation of this parameter is complex and takes into account multiple variables, such as traffic conditions (average delay) on each segment on the planned road adjusted for the time of the day, legal speed limits, road closures and in some cases also weather conditions. A reason for the inaccuracy of the ETA is the lack of consideration for the personal \textit{driving style} of each driver.

\subsection{Business-to-Business}
A common scenario in Business-to-Business consists of two companies that want to benefit from each other's data without having to explicitly share them. The reasoning behind this can be business- or privacy-related. To mitigate risk, both companies allow each other to perform computations using their data in a Federated manner. Many companies are currently exploring such processes in the financial \cite{yang2019federated} or healthcare sectors \cite{chen2019fedhealth}.

In intelligent transportation systems we identified a use-case related to the monetization of connected vehicles. As forecasted by McKinsey, the market value will reach USD 450-750 Billion by 2030 \cite{bertoncello2016monetizing}. This is an opportunity for companies to extend their offering in terms of \textit{analytics as a service} (AaaS) to OEMs. The main obstacle for such a business model is that data is one of the most valuable assets for an OEM, meaning that OEMs are unwilling to share their data. Federated Machine Learning can solve this issue and make analytics as a service a reality, today. In fact, it will grant companies the opportunity to run complex analytics on top of OEMs data without explicitly accessing it.

\section{Product Life Cycle}
The use of Federated Machine Learning will influence the way products are designed, created and shipped to customers \cite{kairouz2019advances}. The whole life cycle will be impacted, especially in such a regulated sector as the automotive industry, where the average time-to-market is high: only 22\% of the players have a go-to- market timeline lower than 13 months\footnote{https://www.jabil.com/blog/automotive-industry-trends-point-to-shorter-product-development-cycles.html}. However, companies like Tesla completely disrupted the market with over-the-air software updates that allow for changes in a car’s functionalities, behaviour and in terms of in-car Machine Learning capabilities\footnote{https://blogs.lse.ac.uk/management/2018/02/05/how-tesla-is-changing-product-life-cycle-in-the-car-industry}. Federated Machine Learning can allow for new possibilities for the automotive industry, thus reducing the overall time-to-market.

Our analysis of how companies can benefit from PET technologies such as Federated Machine Learning lead us to the definition of a new product life cycle for a product in the automotive industry \cite{kairouz2019advances}:
\begin{enumerate}
    \item \textbf{Core software components}: The Federated Machine Learning core is shipped with the software on the end device, e.g., a vehicle or a smartphone with no initial functionalities;
    \item \textbf{Problem identification}: The main actors identify a problem that can be solved using Federated Machine Learning;
    \item \textbf{Client instrumentation}: The end devices are updated to acquire and store data locally. In many cases this data will already be stored independently of the use-case. An example is the location history of the driver;
    \item \textbf{Simulation prototyping (optional)}: Various prototype machine learning models and corresponding architectures are tested to find optimal configuration in a Federated Machine Learning simulation (before shipping them to the end devices);
    \item \textbf{Model training}: Training tasks are started and dispatched (over-the-air) to the set of available end devices. In this configuration we cannot assume that all end devices are available at the same time;
    \item \textbf{Model evaluation}: Trained models are evaluated by computing a set of metrics for the associated tasks till the best model is found. Computing these metrics means either a) computing them after training for each individual end device or b) transmitting the parameters of the trained model to each device and then computing the metrics using the local data;
    \item \textbf{Deployment}: The best model is deployed to each individual end device. Optionally, one could choose to fine-tune the model based on the client’s data as a form of personalisation, to provide a better user experience.
\end{enumerate}

\section{Challenges}
Even though Federated Machine Learning seems like an appealing solution to data-driven decision making without compromising the user privacy and ethics, it still presents some challenges. We grouped them in three categories: \textit{Explainability}, \textit{Availability} and \textit{Security}.

\subsection{Explainability}
Explainability is a common problem in Artificial Intelligence and is one of the key pillars of the ethics guidelines for \textit{Trustworthy Artificial Intelligence}, created by the European Commission\footnote{https://ec.europa.eu/digital-single-market/en/news/ethics-guidelines-trustworthy-ai}. This is specifically true when artificial intelligence is applied to critical systems, such as the automotive sector. It is also called \textit{transparency} and refers to the fact that data, systems and AI business models and decisions should be transparent, or in other words, should be explained in a manner adapted to the stakeholders it concerns. Also, humans need to be aware that they are interacting with an AI system, and must be informed of the system’s capabilities and limitations.

The problem is grounded in the black-box nature of most machine learning techniques, that do not guarantee access to a valuable feedback about what is really happening and why an output has been generated. Some techniques allow to mitigate this challenge and provide an explanation, for example by performing analyses on the existing (centralised) data\footnote{An example is Exploratory Data Analysis, which focuses on summarizing the main characteristics of the dataset.}. However, with Federated Machine Learning this
process cannot be performed, due to the intrinsic characteristic of the (distributed) data. From a customer's perspective, the explanation of the workings of the model can be achieved by elaborating on how a defined input can lead to a particular output. As privacy-awareness grows from a customer's perspective, models need to be carefully designed such that one knows what implications may or may not arise when using it.

\subsection{Availability}
In a Federated Machine Learning setup, where we do not rely on a cloud, we have to consider the situation when devices are \textit{unavailable}. In fact, we rely on end devices, being them smartphones, cars or IoT in general, with limited performances and intermittent connectivity. Such devices can have a considerable downtime for multiple reasons: (1) no connectivity, (2) data caps and (3) battery constraints (for smartphones). Another important factor when considering availability is the number of devices that need to be reached as with Federated Machine Learning this number can easily grow to hundreds of millions. In such a situation, the coordination among the devices needs to be addressed to guarantee availability. For these reasons, it is indeed a challenge to find the requirements that have to be met before making requests to distributed devices.

\subsection{Security}
Federated Machine Learning by itself is purely a method for privacy preservation, since the data is not moved from where it is generated. However, it does not guarantee security as someone would still be able to intercept the communication and obtain access to the user’s personal model parameters. For this reason, other PETs can be combined with Federated Machine Learning to secure the updates and the communication from leakages. An example of adding extra layers of security is the use of homomorphic encryption.

\section{Conclusions and Future Work}
Data is fundamental for decision making and customers want increasingly personalized services. However, with the recent changes in legislations, it is proving to be more difficult to deal with personal data. For this reason, we introduced the topic of Federated Machine Learning as an example of privacy- enhancing technology that can be beneficial for the transportation industry from a privacy and, most importantly, an ethical point of view. We outlined various use-cases that were identified in our company and presented how a product lifecycle will be affected by its adoption, specifically for the automotive industry.

\section*{Ackowledgments}
Thanks go to various colleagues within TomTom for helpful discussions, in particular Łukasz Kosma and Wim Mooij, together with the NAV Research and Innovation and Datalab teams.

\bibliographystyle{unsrt}  
\bibliography{arxiv}

\end{document}